\documentclass[nocite,nobm]{epl}

\begin{document}

\title{The Amplitude of Non-Equilibrium Quantum Interference in Metallic Mesoscopic Systems}
\author{C. Terrier, 
D. Babi\'{c}\footnote{Present address: Department of Physics, Faculty of Science, 
University of Zagreb, Croatia.}, 
C. Strunk\footnote{Present address: 
Institute for Experimental and Applied Physics, 
University of Regensburg, Germany.}, 
T. Nussbaumer and C. Sch\"{o}nenberger}
\institute{
Institute of Physics, University of Basel, Klingelbergstrasse 82, 
CH-4056 Basel, Switzerland 
}

\maketitle

\begin{abstract}

We study the influence of a DC bias voltage $V$ on
quantum interference corrections to the 
measured differential conductance
in metallic mesoscopic wires and rings.
The amplitude of both universal conductance fluctuations (UCF) 
and Aharonov-Bohm effect (ABE) is enhanced
several times for voltages larger than the Thouless energy. 
The enhancement persists even in the presence of
inelastic electron-electron scattering up to 
$V\sim 1$\,mV.
For larger voltages electron-phonon collisions lead to
the amplitude decaying as a power law for the UCF and 
exponentially for the ABE.
We obtain good agreement of the experimental data
with a model which takes into account the decrease
of the electron phase-coherence length due to 
electron-electron and electron-phonon scattering.

\end{abstract}

If the size of a conductor is of order of the
electron phase-coherence length $L_\varphi (T)$ the wave character
of electrons leads to experimentally observable quantum interference contributions to 
the conductance $G$. 
These are the aperiodic and periodic fluctuations $\delta G$ of
the conductance around its average value. In the former case,
called universal conductance fluctuations, the interference pattern is 
formed as a superposition of 
contributions from a continuous range of possible interference paths. 
Averaged over either impurity configuration,
magnetic field or energy, the root-mean-square (rms) conductance 
fluctuation $\delta G_{rms}$ is
of the order $\sim e^2/h$ \cite{leealt}. The periodic fluctuations, known as the
Aharonov-Bohm  effect \cite{ab}, are observed
if the interference is imposed by geometry of a sample, most commonly 
in the form of a loop. 
If the loop is threaded by a magnetic
flux $\phi$,  $\delta G (\phi)$ exhibits periodic oscillations with a 
period $h/e$, and
$\delta G_{rms}$ is again $\sim e^2/h$. 
Both the UCF \cite{skocpol} and the ABE \cite{umbach}
are suppressed by ensemble averaging if
independent phase-coherent units are connected 
in series.
The above behaviour is characteristic of the linear-response regime, 
{\it i.e.} of $eV \ll k_B T$ or $eV \ll E_c$, where $E_c$ is the coherence energy (Thouless energy)
determined by the size of the conductor.
In non-equilibrium ($eV \gg E_c, k_B T$) the fluctuations are expected to 
be remarkably different, as 
predicted theoretically by Larkin and Khmel'nitski\u{\i} (LK) \cite{lk}. 
If inelastic processes can be neglected
the rms fluctuation $\delta g_{rms}$ of the {\it differential conductance} $g$ 
increases with $V$ according to
$\delta g_{rms} \sim (e^2/h) \sqrt{V/V_c}$, where $V_c=E_c/e$. 
This, at first sight surprising result, can be understood as follows.  At $V \gg V_c$
the relevant energy range for the transport subdivides into $N=V/V_c$ 
uncorrelated energy  intervals, each  contributing to the fluctuations of the current by an amount
$\sim (e^2/h) V_c$. Incoherent superposition of these contributions leads to 
$\delta g_{rms}$ being $N^{1/2}$ times larger than $e^2/h$.
Inelastic scattering at large voltages destroys quantum interference.
The enhancement of
$\delta g_{rms}(V)$ is thereby suppressed and 
$\delta g_{rms}$
eventually decreases with increasing voltage \cite{lk}.
Experimental studies  of both the
UCF \cite{webb1,kaplan,ralph,schafer} and ABE \cite{webb1,ct,haus} under
non-equilibrium conditions have been done, but did not result
in a satisfactory understanding of $\delta g (V)$.
Moreover, \cite{ralph} reported on a voltage-independent $\delta g_{rms}$,
\cite{webb1} on a decrease of $\delta g_{rms}(V)$ for $V > V_c$,
whereas in \cite{schafer,ct,haus}  it was found that $\delta g_{rms}(V)$   
shows a non-monotonic behaviour. 
We also note that explanations \cite{ct,haus} of the non-monotonic behaviour of 
$\delta g(V)$, observed in the present work as well, were at a qualitative level.

In this paper we report on the non-equilibrium UCF and ABE in diffusive 
gold samples measured
over a wide voltage range of $V \gg k_B T / e, V_c$ 
(strongly $V$-dominated energy averaging)
that covers 
both the low-voltage $(V/V_c)^{1/2}$ 
enhancement of $\delta g_{rms}$ and its suppression at large voltages.
Emphasis is put on the 
decay of $\delta g_{rms}^{UCF}(V)$ and  $\delta g_{rms}^{ABE}(V)$.
It is shown that $\delta g_{rms}(V)$ decays as a power law for the UCF and 
exponentially for the ABE. A quantitative comparison between simple models and
the experiments allows to extract the voltage-dependent phase-coherence length $L_{\varphi}(V)$ 
and to discuss the nature of inelastic scattering processes out
of equilibrium.

The samples were produced by electron-beam lithography 
and evaporation of 99.99 \% pure gold.
The substrate was silicon covered with 400 nm of SiO$_2$. 
Three 20 nm thick samples of different planar geometries were 
prepared \cite{samples}: (1) 
for the ABE measurements
a ring of average diameter 1 $\mu$m and line width 0.09 $\mu$m (sample S$_{ABE}$),
with a resistance of 25.8 $\Omega$ at 0.3 K; 
(2)  for the UCF measurements a $L_w = 1.5$ $\mu$m long and 0.13 $\mu$m wide wire
(sample S$_{UCF}$), with a resistance of 18.4 $\Omega$ at 0.3 K;
(3)  for the weak-localisation (WL) measurements a  98 $\mu$m long and 0.17 $\mu$m wide wire
(sample S$_{WL}$).
Sample S$_{WL} $ was made relatively long in order to suppress the UCF.
All samples were produced under identical conditions
(the same source of gold and the same evaporation parameters). The diffusion
constants $D$ are consequently very similar:
116 cm$^2$/s for  samples S$_{UCF}$ and S$_{WL}$, and 
110 cm$^2$/s for sample S$_{ABE}$.
The measurements were carried out in a $^3$He cryostat 
with cryogenic rf-filtering, using a low-frequency ($37$\,Hz) lock-in technique 
to measure the differential conductance.
Typical voltage resolution was $\sim 0.3$ nV.
Non-equilibrium UCF and ABE were measured at $T=300$ mK by superimposing
a comparatively large DC bias voltage $V_{DC}$ to the small excitation voltage
$V_{AC}$. The following hierarchy of energies was 
always maintained: $eV_{DC} \gg k_B T \geq eV_{AC} > E_c$. 
Ensemble averaging was achieved by measuring the UCF and ABE over a
magnetic-field range of $\sim 2$\,T, largely exceeding the correlation field.
Typical sweep rates were 0.1 mT/s.
Fig.1a displays raw data of a differential-conductance measurement  on sample S$_{ABE}$,
taken at $V_{DC}=0.5$ mV. Both types of the fluctuations are present. In Fig.1b we show the Fourier
transform of the same data, exhibiting a well-defined peak at the position corresponding to
$\phi=h/e$. 
From the magnetoconductance traces we have extracted 
$\delta g_{rms}^{ABE}$ and $\delta g_{rms}^{UCF}$ for sample S$_{ABE}$, as shown
in Fig.2 (discussed in more detail later), and $\delta g_{rms}^{UCF}$ for sample S$_{UCF}$.
The values of $\delta g_{rms}^{UCF}$ have been determined as standard deviations of the 
whole differential-conductance traces. The amplitudes  $\delta g_{rms}^{ABE}$ have been calculated by 
averaging the periodic part of the fluctuations over  $\sim 500$ periods, after the coarse background 
has been removed numerically.

The WL was measured on sample S$_{WL}$ in the linear-response regime ($V_{DC}=0$) and in the 
temperature range 0.3 - 10 K.
By fitting a one-dimensional WL expression \cite{altshulerwl} to 
the low-field magnetoconductance data 
we obtained the linear-response $L_\varphi$ as a function of temperature, as shown in the inset to Fig.3.   
Similarly to published work \cite{mohanty,gougam},
$L_\varphi (T)$ saturates at low temperatures (below 1 K), 
which has been attributed to impurity mediated 
inelastic electron-electron scattering \cite{glazman}.
At high temperatures $L_\varphi$
follows a $L_\varphi \propto T^{-q}$ dependence with $q \approx 1.2$ that 
suggests dephasing by electron-phonon interaction.
At $T= 300$ mK, $L_\varphi = 3$ $\mu$m and 
$\tau_\varphi = \sqrt{L_\varphi^2 / D} =  0.77$ ns,
implying that in the linear-response regime the diffusion of electrons is coherent throughout the samples
$S_{UCF}$ and $S_{ABE}$.
$L_\varphi$ has to be compared further with the
effective length of the sample $L_{c0}$ which depends on the particular geometry 
and the coherence phenomenon investigated. For the UCF in sample S$_{UCF}$,
$L_{c0} = L_w = 1.5$ $\mu$m. In the case of ABE in sample S$_{ABE}$,
$L_{c0} = C_r \approx 3.14$ $\mu$m, where $C_r$ is the circumference of the ring.
Finally, for the UCF in sample S$_{ABE}$, $L_{c0}$
is calculated as  the distances ring - voltage contacts plus $C_r /2$, which
gives $\approx 2.2$ $\mu$m. 
Thus, $L_\varphi$(300 mK) is always bigger or very similar to 
$L_{c0}$, so that the linear-response coherence
energy is set by the time of electron diffusion through the sample
according to $E_{c0} = eV_{c0} = \hbar D / L_{c0}^2$.

With increasing temperature or applied voltage the coherence length decreases,
and once it becomes shorter than $L_{c0}$ the interference
contributions are suppressed. 
The effect
of temperature was demonstrated by Milliken {\it et al.} \cite{milliken}.
Their method was to determine  $L_\varphi (T)$ from the WL and to use this to
describe the decay of {\it equilibrium} $\delta G$, which turned out to be a power 
law for the UCF and exponential for the ABE. Our approach is complementary: 
we keep the bath temperature constant and investigate the voltage dependence
of the {\it non-equilibrium}
$\delta g$.  While Milliken {\it et al.} observed
a monotonically-decreasing $\delta G (T)$, in our case $\delta g (V)$ is
a non-monotonic function, as shown  in Fig.2.
For small $V_{DC}$ (but still much larger than $V_{c0}$), 
$\delta g_{rms}$ is enhanced by $V_{DC}$. At higher voltages
$\delta g_{rms}$ decreases with $V_{DC}$, {\it faster for the} ABE
{\it than for the} UCF,
which is in qualitative agreement with the results of Milliken {\it et al.}

Let us first turn to the details of non-equilibrium UCF. For convenience we
denote the phase-coherence length at a finite bias voltage by $L_{\varphi}(V)$, while
$L_{\varphi}(T)$ referes to the linear-response value. 
If $V_{DC} \gg V_{c0}, k_B T/e$ and
$L_{\varphi} (V_{DC}) > L_{c0}$, the LK enhancement of $\delta g_{rms}$ holds and one
expects that $\delta g_{rms} \propto \sqrt{V_{DC}/V_{c0}}$. 
Once $L_{\varphi} (V_{DC})$  becomes substantially
smaller than $L_{c0}$ inelastic processes start to be important for most of 
the electrons, and the sample effectively 
subdivides into $L_{c0}/L_ {\varphi}$ uncorrelated phase-coherent  sections in series. 
The LK enhancement in a single section is still valid if the voltage
drop $V^{sec}_{DC} = V_{DC} (L_{\varphi}/L_{c0})$  over a section exceeds the 
corresponding coherence voltage $V_c^{sec} = V_{c0} (L_{c0}/L_{\varphi})^2$.
Following the LK arguments, $\delta g_{rms}$ corresponding to a single
section is proportional to 
$(V^{sec}_{DC} / V_c^{sec})^{1/2} = [ (V_{DC}/V_{c0}) (L_{\varphi}/L_{c0})^3]^{1/2}$.
Incoherent addition of these contributions of sections in series 
leads to the {\it differential conductance} being suppressed by a factor
$(L_{\varphi}/L_{c0})^{1/2}$, and hence
\begin{equation}
\delta g_{rms}^{UCF} = A_{UCF} \frac{e^2}{h}     \sqrt{ \frac{V_{DC}}{V_{c0}} } 
\left( \frac{L_{\varphi} (V_{DC})}{ L_{c0}} \right) ^2 \; ,
\end{equation}
where $A_{UCF}$ is a constant prefactor. This expression is 
different from that resulting from the linear-response ensemble averaging, 
and valid if $V_{DC} > V_{c0}$ and $L_{\varphi} < L_{c0}$ \cite{remark}. It enables an
unambiguous determination of $L_{\varphi} (V_{DC})$. 
Since the form of $L_{\varphi} (V_{DC})$ is not known {\it a priori}, 
we cannot determine $A_{UCF}$ and $L_{\varphi} (V_{DC})$
simultaneously from a direct fit to the data. 
To avoid this problem we fix $A_{UCF}$  
by setting $L_\varphi= L_{c0}$
for the first measured point which satisfies $V_{DC} \gg V_{c0}$,
and then simply extract
$L_{\varphi} (V_{DC})$ for all the other points from the measured data. 
For the two sets of UCF data, {\it i.e.} for samples S$_{UCF}$ and
S$_{ABE}$, we have obtained excellent agreement in $L_{\varphi}(V_{DC})$, 
as we show in Fig.3 by open 
(S$_{ABE}$) and full (S$_{UCF}$) squares. Normalised to the 
corresponding  values of $L_{c0}$
both results collapse onto a single curve. 
The phase-coherence length deduced from our non-equilibrium
experiment $L_{\varphi}(V)$ has a qualitatively similar functional form
as $L_{\varphi}(T)$ obtained from the WL measurements
(apart from the saturation of $L_\varphi (T)$, since
saturation of $L_\varphi (V)$ is not expected 
for $eV \gg k_B T$ \cite{glazman}). Two regimes
with different power-law dependences are discernible.
In the region of increasing $\delta g_{rms}(V_{DC})$ there is a power law 
$L_{\varphi}(V_{DC}) \propto V_{DC}^{-s}$, where $s = 0.19 \pm 0.02$. 
For decreasing $\delta g_{rms} (V_{DC})$, {\it i.e.} where $L_{\varphi}(V_{DC})$ 
is considerably smaller than $L_{c0}$,
the dependence changes to $L_{\varphi} \propto V_{DC}^{-p}$ with 
$p = 0.57 \pm 0.03$. Inserting thus found $L_{\varphi}$ back into Eq.1 results in 
the dotted line in Fig.2.

We now discuss the non-equilibrium ABE. Here the LK theory cannot be 
directly applied to extract $\delta g_{rms}$. 
Subdivision of a ring into phase-coherent sections makes no sense in this case, as only those
electrons which stay coherent over the whole length $C_r$ contribute to the interference. However, since the
same physics governs the UCF and ABE, 
$\delta g_{rms}^{ABE}$ is expected to be $\propto \sqrt{V_{DC}/V_{c0}}$
in the regime where inelastic scattering is absent.
This is indeed the case, as shown in Fig.2.
The form of the suppression of  $\delta g_{rms}^{ABE}$ can be inferred from a theoretical analysis of
DiVincenzo and Kane \cite{dk}. They have shown that inelastic processes influence the ABE in two ways.
First, the fluctuations are suppressed as $\exp( - \beta C_r / L_{\varphi})$, where $\beta$ is 
of order unity in the range $1 < C_r / L_{\varphi} < 3$ 
and smaller in the case of a true exponential decay occurring in
the asymptotic limit $C_r / L_{\varphi} \rightarrow \infty$. Second, the proper form of $E_c$ in the presence
of inelastic processes is given by 
$E_c^{in} = \hbar D / C_r^{2-\alpha} L_{\varphi}^\alpha = E_{c0} (C_r/L_{\varphi})^\alpha$ 
with $\alpha \approx 1.3$. This renormalisation of the coherence energy is a consequence of the statistics
for those electrons that diffuse around a ring without being scattered inelastically \cite{dk}.
Roughly, the probability distribution for an electron to contribute to the ABE has a maximum
at a value of the electron traversal time $\sim D /C_r L_{\varphi}$ \cite{dk}.
Combining the above arguments we can write for $L_{\varphi} < C_r$:
\begin{equation}
\delta g_{rms}^{ABE} = A_{ABE} \frac{e^2}{h}     \sqrt{ \frac{V_{DC}}{V_{c0}} } 
\left(  \frac{L_{\varphi}(V_{DC})}{C_r} \right) ^{0.65} 
e^ { - \beta C_r / L_{\varphi}(V_{DC}) } \; .
\end{equation} 
Good agreement of the ABE data with both Eq.2 and the UCF data 
is obtained for $\beta \approx 1.1$, 
as we show in Fig.2 by the solid curve and in Fig.3 by crosses.

Below we discuss how the observed power law dependences
in $L_\varphi (V_{DC})$ can be linked to microscopic scattering processes.
From Figs.2,3 we can distinguish two 
regimes (denoted by {\rm I} and {\rm II} in Fig.3): 
{\rm (I)} $\delta g_{rms} (V_{DC})$ increases with  $V_{DC}$ and
$L_{\varphi}(V_{DC}) \propto V_{DC}^{-0.19}$, and {\rm (II)},
$\delta g_{rms}(V_{DC})$ decreases with $V_{DC}$ while
$L_{\varphi}(V_{DC}) \propto V_{DC}^{-0.57}$. 

The phase-coherence length is determined by the dephasing rate
$\tau_{\varphi}^{-1}$ via $L_{\varphi}=\sqrt{D\tau_{\varphi}}$.
$\tau_{\varphi}^{-1}$ can be expressed as an average of the
inelastic scattering rate $\tau_{in}^{-1}(\epsilon)$, ($\epsilon$
is the energy exchanged in the interaction) of electrons over the
accessible energy range given by the width of the electron
distribution function $f$ \cite{gougam}. If the electron-electron
scattering is dominant, the scattering rate is energy dependent as
described by the kernel function
$K(\epsilon)=\kappa_{\eta}\epsilon^{-\eta}$. The energy interval
$E$ accessible for the scattering is set by $k_BT$ in equilibrium
and $eV_{DC}$ in non-equilibrium. At high energies, where
single scattering events determine $\tau_\varphi$, a simple
argument \cite{gougam} leads to $\tau_{\varphi,ee}^{-1}(E) \propto
E^{1/\eta}$ \cite{saturation}. Two choices for $\eta$ are currently under debate:
$\eta=3/2$ for the disorder enhanced Coulomb interaction \cite{gougam} and
$\eta=2$  for magnetic impurity mediated interaction
\cite{glazman,pierre}. This results in
$L_{\varphi,ee}^{-1}(V_{DC}) \propto
V_{DC}^{1/3}$ for $\eta = 3/2$, and $L_{\varphi,ee}^{-1}(V_{DC}) \propto V_{DC}^{1/4}$ 
for $\eta = 2$.
Hence the kernel exponent $\eta=2$ appears more consistent with
our experimental observation of $L_\varphi^{-1} \propto
V_{DC}^{0.19}$. This is supported by the observed saturation of
the equilibrium $\tau_\varphi$ for temperatures below \mbox{1 K}.

We emphasise that - independently of the precise interaction
mechanism - the rather weak decrease of $L_{\varphi,ee}$ with
increasing $V$ is insufficient to suppress the LK enhancement of
$\delta g_{rms}$. This is seen as follows. We recall that the
condition for the appearance of the LK enhancement in a
phase-coherent section of the wire is $V_{DC}^{sec} > V_c^{sec}$, which can be written as
$V_{DC}/V_{c0} > (L_{c0}/L_{\varphi})^3$. Since $L_{\varphi,ee}
\propto V_{DC}^{-1/4}$, the right-hand side of the above
inequality increases more slowly with $V_{DC}$ than the left-hand
side, and the condition for the enhancement is maintained over
the whole voltage range of increasing $\delta g_{rms}  (V_{DC})$.

While the regime {\rm I} is determined by electron-electron scattering
and a non-equilibrium electron distribution function $f$, 
the stronger decay of $L_{\varphi}(V)$ in the regime {\rm II} 
is caused by electron-phonon scattering. It can be very well
described by equilibrium properties if we assume local
thermal equilibrium with an elevated electron temperature
$T_{el} (V_{DC})$.
Electron-phonon scattering leads to 
$\tau_{\varphi,ep} \propto T_e^{-m}$, where $m$ equals 3 for very clean samples,
4 for strongly disordered samples, and 2 - 3 for samples of intermediate degree of 
disorder \cite{schmid}.
Since $T_e$ is related to the applied voltage by
$T_e \propto V^{2/(2+m)}$ \cite{anderson}, one obtains 
$\tau_{\varphi,ep} \propto V^{-2m/(2+m)}$ 
and $L_{\varphi,ep} \propto V^{-m/(2+m)}$. Our result 
$L_{\varphi}(V_{DC}) \propto V_{DC}^{-0.57 \pm 0.03}$ gives
$m = 2.6 \pm 0.3$, which agrees well with the value $2.5$ obtained
in a noise measurement on similar samples \cite{henny3}.
Moreover, our WL result
($L_{\varphi,ep}(T)\propto T^{-m/2}$) is $m = 2.4$,
in excellent agreement
with the above values.
We also note that even though $T_e$
rises up to $\sim 10 - 15$ K at high $V_{DC}$ the energy
averaging given by the LK approach remains essentially 
unaffected because $T_e$ increases 
more slowly than the voltage itself. 
Although our phenomenological analysis leads to a
consistent interpretation of our data,
further theoretical work is needed to fully understand
the dephasing in the case of strong non-equilibrium.

In conclusion, we have measured the universal conductance fluctuations 
and Aharonov-Bohm effect in mesoscopic gold
samples under highly non-equilibrium conditions of large applied bias voltages.
The rms fluctuation $\delta g_{rms}$ of 
the differential conductance initially increases with voltage
$\propto \sqrt{V/V_c}$,
which demonstrates the validity of the theoretical 
prediction by Larkin and Khmel'nitski\u{\i}.
This increase is followed by a decay of $\delta g_{rms}$
at higher voltages, where inelastic scattering becomes 
substantial. The amplitude decays as a power law for the universal conductance
fluctuations and exponentially in the case of the Aharonov-Bohm
effect. The decrease of the phase-coherence length with
increasing voltage is in 
good agreement with the inferred $L_{\varphi}(V)$
dependences for electron-electron and 
electron-phonon scattering.
In particular, the electron-electron collisions
are not sufficient to suppress the enhancement mechanism.
This work was supported by the Swiss National Science Foundation.

%
% FIG1
%
\begin{figure}
\onefigure[width=8cm]{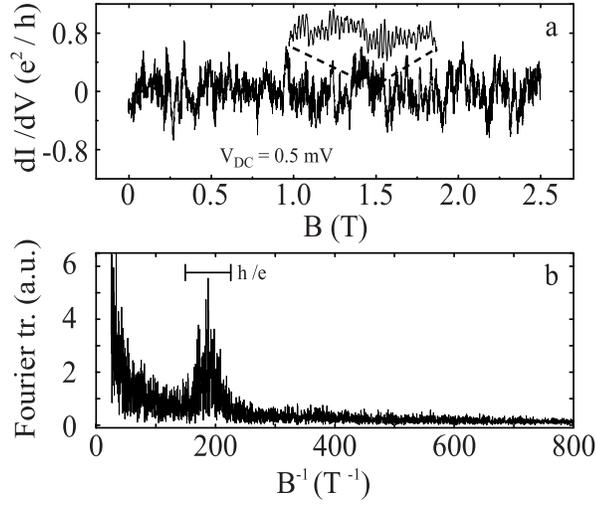}
\caption{(a) A differential-magnetoconductance trace of sample S$_{ABE}$, taken at 
$V_{DC} = 0.5$ mV. 
Both the periodic (expanded view) and aperiodic fluctuations are present. 
(b) Fourier
transform of (a), showing a well-defined peak at the position corresponding to
$\phi = h/e$.}
\label{f.1}
\end{figure}

%
% FIG 2
%
\begin{figure}
\onefigure[width=8cm]{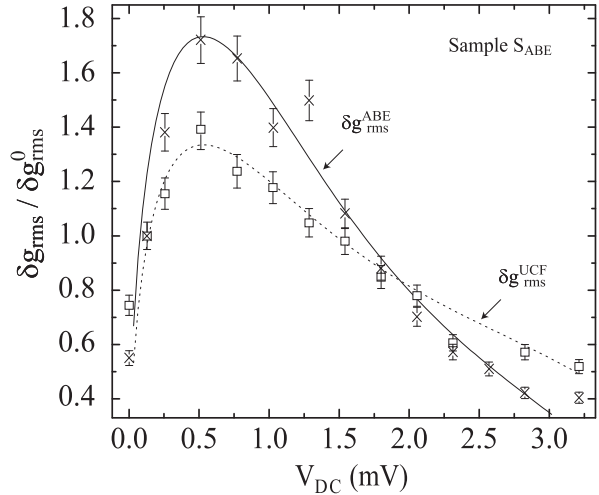}
\caption{Plot of $\delta g_{rms}^{ABE}$ (crosses) and $\delta g_{rms}^{UCF}$ (open squares)
{\it vs.} $V_{DC}$ for sample S$_{ABE}$,
with $\delta g_{rms}$ normalised to the 
values $\delta g_{rms}^0$ at the first $V_{DC} \neq 0$ points (which also satisfy the criterion
$V_{DC} \gg V_{c0}$). The lines are plots of Eq.1 (dashed line)
and Eq.2 (solid line) discussed in the text.}
\label{f.2}
\end{figure}
%
%FIG3
%
\begin{figure}
\onefigure[width=8cm]{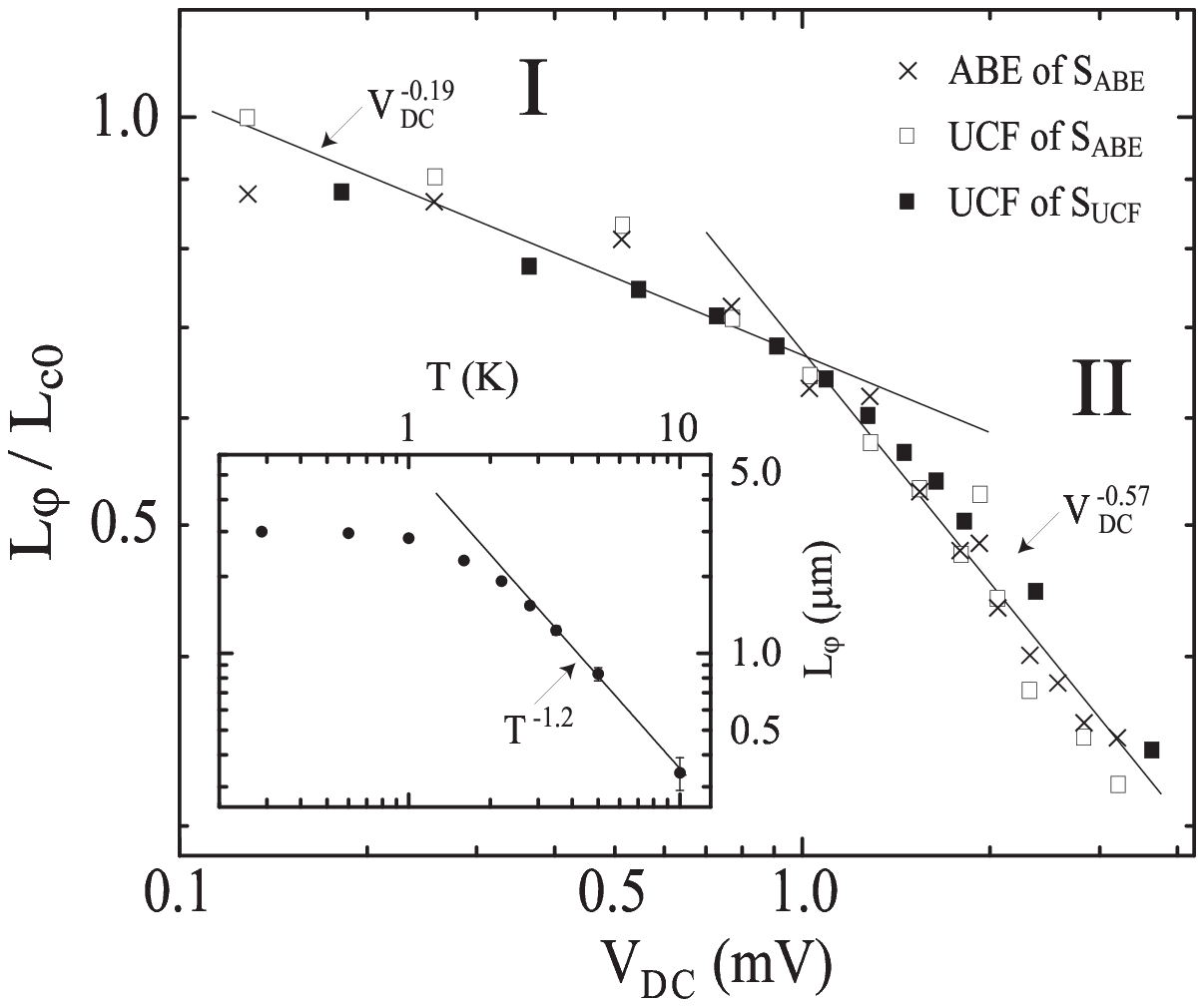}
\caption{Phase-coherence length $L_{\varphi}(V_{DC})$, 
normalised to the characteristic lengths $L_{c0}$
defined in the text,
deduced from the non-equilibrium UCF and ABE measurements.
Crosses: ABE in sample S$_{ABE}$.
Open squares: UCF in sample S$_{ABE}$. Full squares: UCF in sample S$_{ABE}$.
Below $V_{DC} \sim 1$ mV, $L_{\varphi} \propto V_{DC}^{-0.19 \pm 0.02}$, and
above $L_{\varphi} \propto V_{DC}^{-0.57 \pm 0.03}$.
Inset:  Linear-response coherence length $L_\varphi (T)$ found from the WL measurements on
sample S$_{WL}$.}
\label{f.3}
\end{figure}


\begin{thebibliography}{99}

\bibitem{leealt}
  \Name{Lee P. A.\and  Stone D.}
  \REVIEW{Phys. Rev. Lett}{55}{1985}{1662}.

\bibitem{ab}
  \Name{Aharonov Y. \and Bohm D.}
  \REVIEW{Phys. Rev.}{115}{1959}{485}.

\bibitem{skocpol}
  \Name{Skocpol W. J., Mankiewich P. M., Howard R. E., Jackel L. D.,
Tennant D. M. \and Stone A. D.}
  \REVIEW{Phys. Rev. Lett.}{56}{1986}{2865}.

\bibitem{umbach}
  \Name{Umbach C. P., Van Haesendonck C.,  Laibowitz R. B., 
Washburn S., \and Webb R. A.}
  \REVIEW{Phys. Rev. Lett.}{56}{1986}{386}.

\bibitem{lk}
  \Name{Larkin A. I. \and Khmel'nitski\u{\i} D. E.}
  \REVIEW{Pis'ma Zh. Eksp. Teor. Fiz.}{91}{1986}{1815}
  [\REVIEW{Sov. Phys. JETP}{64}{1986}{1075}];
  \REVIEW{ Phys. Scr.}{T14}{1986}{4}.

\bibitem{webb1}
  \Name{Webb R. A. ,Washburn S. \and Umbach C. P.}
  \REVIEW{Phys. Rev. B}{37}{1988}{8455}.

\bibitem{kaplan}
  \Name{Kaplan S. B.}
  \REVIEW{Phys. Rev. B}{38}{1988}{7558}.

\bibitem{ralph}
  \Name{Ralph D. C., Ralls K. S. \and Buhrman R. A.}
  \REVIEW{Phys. Rev. Lett.}{70}{1993}{986}.

\bibitem{schafer}
  \Name{Sch\"{a}fer R., Hecker K., Hegger H. \and  Langheinreich W.}
  \REVIEW{Phys. Rev. B}{53}{1996}{15964}.

\bibitem{ct}
  \Name{Terrier C., Strunk C., Nussbaumer T., Babi\'{c} D. \and Sch\"{o}nenberger C.}
  \REVIEW{Fizika A (Zagreb)}{8}{1999}{157}.

\bibitem{haus}
  \Name{H\"{a}ussler R., Weber H. B. \and  von L\"{o}hneysen H.}
  \REVIEW{J. Low Temp. Phys}{118}{2000}{467}.

\bibitem{samples}
The results for these samples are representative of the data obtained by measurements 
carried out on altogether eleven samples. We did not find any significant sample dependence of the results.

\bibitem{altshulerwl}
  \Name{Al'tshuler B. L.  \and  Aronov A. G.}
  \REVIEW{Pis'ma Zh. Teor. Eksp. Teor. Fiz.}{33}{1981}{515}
[\REVIEW{JETP Lett}{33}{1981}{499}].

\bibitem{mohanty}
  \Name{Mohanty P.}
  \REVIEW{Physica B}{280}{2000}{446}.

\bibitem{gougam}
  \Name{Gougam A. B.,Pierre F., Pothier H., Esteve D. \and Birge N. O.}
  \REVIEW{J. Low Temp. Phys.}{118}{2000}{447}.

\bibitem{glazman}
  \Name{Kaminski A. \and Glazman L. I.}
  \REVIEW{Phys. Rev. Lett.}{86}{2001}{2400}.

\bibitem{milliken}
  \Name{Milliken F. P., Washburn S., Umbach C. P., Laibowitz R. B. \and Webb R. A.}
  \REVIEW{Phys. Rev. B}{36}{1987}{4465}.

\bibitem{remark}
Strictly speaking, the conditions are $V_{DC}\gg V_{c0}$ and $L_{\varphi}\ll L_{c0}$.

\bibitem{dk}
  \Name{DiVincenzo D. P. \and  Kane C. L.}
  \REVIEW{Phys. Rev. B}{38}{1988}{3006}.  

\bibitem{saturation}
The saturation of $\tau_\varphi (T)$ at very low energies is not covered by the simple
argument of Ref. 15.

\bibitem{pierre} 
\Name{ Pierre F., Pothier H., Esteve D., Devoret M. H., Gougam A. B. \and Birge N. O.}
in \Book {Proceedings of the NATO Advanced Research Workshop on Size-Dependent
Magnetic Scattering, Pecs, Hungary}
\Publ{Kluwer, Dodrecht, The Netherlands}
\Year{2001}

\bibitem{schmid}
  \Name{Schmid A.}
  in \Book{Localization, Interaction and Transport Phenomena}
  \Editor{Kramer B., Bergmann G. \and Bruynsereade Y.}
  %\Vol{9}
  \Publ{Springer, Berlin}
  \Year{1985}
%  \Page{666}.

\bibitem{anderson}
  \Name{Anderson P. W., Abrahams E. \and Ramakrishnan T. V.}
  \REVIEW{Phys. Rev. Lett.}{43}{1979}{718}.

\bibitem{henny3}
  \Name{Henny M.} PhD Thesis, University of Basel, 1998. 




\end{thebibliography}
\end{document}